\documentclass[aps,pre,reprint,notitlepage,twocolumn,showpacs,showkeys,groupedaddress,superscriptaddress,amsmath,amssymb]{revtex4-1}
\usepackage{graphicx,subfigure}
\usepackage{epic}
\usepackage{bm}
\usepackage{bbold}
\usepackage{natbib}
\usepackage{hyperref}
\usepackage{color}
\makeatletter

\begin{document}

\title{ Cell motility as an energy minimization process}

\author{H. Chelly and P. Recho}

\date{\today}

\begin{abstract}
The dynamics of active matter driven by interacting molecular motors has a non-potential structure at the local scale. However, we show that there exists a quasi-potential effectively describing the collective self-organization of the motors propelling a cell at a continuum active gel level. Such a model allows us to understand cell motility as an active phase transition problem between the static and motile steady state configurations that minimize the quasi-potential. In particular both configurations can coexist in a metastable fashion and a small stochastic disorder in the gel is sufficient to trigger an intermittent cell dynamics where either static or motile phases are more probable, depending on which state is the global minimum of the quasi-potential.

\end{abstract}

\maketitle

\section{Introduction}

In three-dimensional biological matrices, cell migration usually does not rely on the formation of focal adhesions \cite{paluch2016focal} and, taking advantage the external confinement, uses the non-specific
friction between the cell and its environement \cite{bergert2015force} to exert traction forces that break the system symmetry and lead to motion. Depending on the force production mechanism of the traction forces, several physical models have been put forward to shed light on this instability setting the onset of motility \cite{Ziebert2012,tjhung2012spontaneous,recho2013contraction,CallanJones2013,camley2013periodic,blanch2013spontaneous,barnhart2015balance,giomi2014spontaneous}. In such models, the interaction with the substrate  is  present in the form of a friction coefficient that can be modulated depending on the affinity of the cell and its environment. 

Recently, several two or three dimensional models have been put forward to show that the limit of a vanishing friction coefficient where the \textcolor{black}{power exerted by the traction forces on the substrate is negligible compared to other sources of bulk dissipation}, can still lead to cell motion \cite{loisy2019tractionless,farutin2019crawling,le2020actomyosin}. In such limit, motility \textcolor{black}{ is possible because of the turnover property of the cell skeleton which can build up through polymerization in the vicinity of the leading edge and depolymerize in sinks while the  building blocks requiered to do so are not connected to the substrate \cite{Juelicher2007}. The cell material is then continuously renewed ahead of the cell front and can support a traction-free motion.    Interestingly, in such paradigmatic situation, motility} becomes an intrinsic property of the cell that is independent of the environment biophysical details.   One can also speculate on the biological role of such mechanism as it would render cell motion robust with respect change of the environment chemistry and rheology.

Assuming that cell propulsion in a confined environment such as a track or a channel \cite{maiuri2012first,doyle2013dimensions} is mainly driven by its molecular motors \cite{paluch2016focal}, we study one of the most simple one-dimensional model of this substrate independent type of cell motility. We show that, despite its active nature, our model has a variational structure  with an effective quasi-potential that is minimized in the course of the cell motion and that the minima of the quasi-potential correspond to the model metastable steady states. These minima represent  a static symmetric configuration or a motile asymmetric configuration of the cell and their appearance and relative level is controlled by two non-dimensional parameters driving the motors self-organization: a global contractility coefficient and a parameter representing the steric hindrance between the motors.  

Next, by introducing a small stochastic perturbation in the active stress, we show that the metastability  of the deterministic system leads to intermittent cell dynamics which can be either dominated by static phases or by motile phases depending on which state is the global or local minimum of the quasi-potential. \textcolor{black}{Although our minimal model aims at establishing a physical paradigm rather than reproducing some specific experimental data, this result may have importance to physically rationalize some experimentally observed phenomena such as the intermitency of individual cell dynamics \cite{maiuri2015actin,hennig2020stick} or the fact that in a population of similar cells, a proportion is motile while others are static \cite{kwon2019stochastic}.}

\section{Contraction driven motion}

A simple physical paradigm describing contraction-diven cell motility on a stiff substrate is presented in  \cite{recho2013contraction,Recho2015}. In this model  the cell skeleton can be represented as a segment with a fixed length \textcolor{black}{moving} on a one-dimensional track. More generally, for a deformable substrate \cite{Wong2011}, the stress balance in the skeleton reads 
\begin{equation}\label{e:f_bal_1}
\partial_x\sigma=\xi(v-v_s),
\end{equation} 
where $x\in [l_-(t),l_+(t)]$ is the spatial coordinate labeling material points of the cell skeleton,  $t>0$ is the time, $l_-(t)$ and $l_+(t)$ are the moving fronts of the cell, $\sigma(x,t)$ is the axial stress, $\xi$ is a friction coefficient, $v(x,t)$ is the velocity of the skeleton and $v_s(x,t)$ is the velocity of the substrate. Supposing that the two moving fronts are connected by a stiff spring representing the cell volume regulation mechanism \cite{putelat2018mechanical}, we can associate the following boundary conditions to \eqref{e:f_bal_1}:
\begin{equation}\label{e:f_bal_bc_1}
\sigma(l_-(t),t)=\sigma(l_+(t),t) \text{ and } L=l_+(t)-l_-(t),
\end{equation} 
where $L>0$ is the fixed cell length. Since the incoming fluxes of skeleton at the cell boundaries vanish, we have:
\begin{equation}\label{e:bc_Stefan}
V(t)\stackrel{\text{def}}{=}\partial_tl_{-}(t)=\partial_tl_{+}(t)=v(l_{-}(t),t)=v(l_{+}(t),t),
\end{equation}
where $V$ is the velocity of the cell. The skeleton constitutive behavior is assumed to be that of a visco-contractile active gel \cite{Juelicher2007},
\begin{equation}\label{e:const_be}
\sigma=\eta \partial_xv+\chi c,
\end{equation}  
where $\eta$ is the skeleton \textcolor{black}{viscosity}, $\chi$ is the motor contractility and $c(x,t)$ is the concentration of motors cross-linking the skeleton filaments. \textcolor{black}{Notice that this simple description only models the contraction-driven skeleton flow setting the cell fronts velocity. Although the skeleton building blocks polymerization and depolymerization is not described as this process follows the skeleton flow without impacting it in our perspective (see Appendix~\ref{sec:appendix_0}), such turnover  is nonetheless essential to reconstruct a realistic skeleton density \cite{recho2013contraction}.} Following Appendix~\ref{sec:appendix_A}, we assume that the motor concentration follows the non-linear drift-diffusion equation
\begin{equation}\label{e:motors_dd_1}
\partial_tc+\partial_x(cv-D\partial_x(f(c/c_0)c))=0,
\end{equation} 
where $D$ is an effective diffusion coefficient, $f$ is a non-dimensional positive and non-decreasing function that accounts for the inhibition of the motors attachment to the skeleton at a high concentration due to a \textcolor{black}{steric hindrance constraint \cite{truong2021extent}} and 
\begin{equation}\label{e:total_mass_1}
c_0=\frac{1}{L}\int_{l_-}^{l_+}c(x,t)\mathrm{d}x.
\end{equation}
is the average concentration of motors. Because the fluxes of motors through the cell boundaries vanish
\begin{equation}
\partial_xc(l_{\pm}(t),t)=0,
\end{equation}
$c_0$ is a constant set by the initial concentration.

Finally, the substrate is assumed to be visco-elastic so that certain functional $\mathcal{L}$ relates its velocity with the traction forces exerted by the cell,
$v_s=\mathcal{L}[\partial_x\sigma].$
Clearly, if the traction forces $\partial_x\sigma$ vanish, the substrate velocity is also zero: $\mathcal{L}[0]=0$.

\section{Substrate independent regime}
\textcolor{black}{In this paper, we consider the case of a  vanishing friction coefficient, $\xi\rightarrow 0$. This limit physically means that the dissipation due to the interaction with the substrate is negligible compared to the bulk viscous dissipation. More specifically, combining \eqref{e:f_bal_1} and \eqref{e:const_be} with boundary conditions \eqref{e:f_bal_bc_1} and \eqref{e:bc_Stefan}, we obtain the following balance of powers \cite{PhysRevLett.112.218101}:
$$-\chi\int_{l_-}^{l_+}c\partial_xv \mathrm{d}x=\eta\int_{l_-}^{l_+}(\partial_xv)^2\mathrm{d}x+\xi\int_{l_-}^{l_+}(v-v_s)v\mathrm{d}x.$$
The lefthandside of the above relation is the active power performed by the molecular motors to deform the cell skeleton meshwork. It is dissipated at the righthandside by the skeleton viscosity and its interaction with the substrate which can itself  be decomposed into the dissipation due to the relative frictional velocity and the visco-elastic dissipation in the substrate bulk. Denoting $\bar{v}$ the typical scale of velocities, in the regime that we consider, we therefore have the scaling relations
$$\chi c_0\bar{v}/L\sim \eta (\bar{v}/L)^2 \text{ and } \xi \bar{v}^2 \ll\eta (\bar{v}/L)^2.$$
Thus $\bar{v}\sim \chi c_0 L/\eta$ and $L\ll \sqrt{\eta/\xi}$ and the vanishing friction limit corresponds to the situation where the cell length is much smaller than the hydrodynamic length $\sqrt{\eta/\xi}$ screening the stress propagation in the skeleton \cite{saha2016determining}. In this situation, the propagation of the stress locally created by a bundle of molecular motors is long-range as it spans over the whole skeleton meshwork. This approximation is not directly applicable to the well-characterized case of fish keratocytes crawling on a two dimensional surface for which it can be roughly estimated that $\eta\simeq 10^5$ Pa s and $\xi\simeq 10^{16} \text{ Pa m}^{-2}\text{ s}$ \cite{Barnhart2011}, rather leading to $\sqrt{\eta/\xi}\sim L$. But we anticipate that this limit, aside from its conceptual interest, can be important for other cells types that move in the bulk of an extra-cellular matrix \cite{even2005cell} where the adhesion with the environment is usually weaker.  }

\textcolor{black}{When the friction with respect to the substrate can be neglected compared to the internal friction represented by viscosity, as the skeleton and substrate velocities remain bounded, we locally have $\partial_x\sigma\simeq 0$ in  \eqref{e:f_bal_1}, leading to  $v_s=0$. In the case where $\xi=0$, the mechanical problem is ill-posed as any arbitrary rigid body motion can be superimposed to the movement.  However, from the boundary conditions \eqref{e:f_bal_bc_1} imposing the same stress at the two fronts, we obtain the global constraint 
$$\xi\int_{l_-}^{l_+}(v-v_s)\mathrm{d}x=0,$$
which we use, supposing that $\xi$ is not exactly zero,  to impose the condition:
$$\int_{l_-}^{l_+}v\mathrm{d}x=0.$$
Such global constraint is sufficient to eliminate the rigid body motions and define unambiguously the vanishing friction limit which leads to a generic cell  dynamics that is independent of the cell/substrate mechanical behavior. }

\section{Model formulation}

Combining the constitutive relation \eqref{e:const_be} with the no-flux boundary conditions \eqref{e:bc_Stefan}, we obtain that the homogeneous stress in the skeleton is $\sigma=\chi c_0$. As a result, $\chi(c_0-c)=\eta\partial_xv$ which leads by integration to,
$$v(x,t)-V(t)=\frac{\chi}{\eta}\int_{l_-}^{l_+}\text{H}(x-z)(c_0-c(z,t))\mathrm{d}z,$$
where $\text{H}$ denotes the Heaviside step function.

Defining the non-dimensional traveling coordinate $y=[x-(l_-+l_+)/2]/L$ and rescaling the concentration by $c_0$, the space by $L$ and the time by $L^2/D$, we obtain the following non-dimensional coupled problem:
\begin{equation}\label{e:no_frict_motil_pb}
\left\lbrace \begin{array}{c}
\alpha(1-c)= \partial_yw\\
\partial_tc+\partial_y(cw-\partial_y(f(c)c))=0,
\end{array}\right. 
\end{equation}
with no-flux boundary conditions on $c$, $\partial_yc(\pm 1/2,t)=0$ and $w$, $w(\pm 1/2,t)=0$. In \eqref{e:no_frict_motil_pb}, there is a single non-dimensional parameter  $\alpha=\chi c_0 L^2/(\eta D)$ sets the importance of the contractile activity compared to the two dissipative mechanisms of diffusion and viscosity. As $w=v-V$ represents the flow of skeleton in the cell frame of reference, the cell velocity is given by the condition,
\begin{equation}\label{e:cell_velocity}
V(t)=-\int_{-1/2}^{1/2}w(y,t)\mathrm{d}y. 
\end{equation}
System \eqref{e:no_frict_motil_pb} can also be written as a single non-linear and non-local drift-diffusion equation by solving for $w$ in $\eqref{e:no_frict_motil_pb}_1$, 
\begin{equation}\label{e:w_c}
w(y,t)=\alpha \int_{-1/2}^{1/2}\text{H}(y-z)(1-c(z,t))\mathrm{d}z
\end{equation}
such that $\eqref{e:no_frict_motil_pb}_2$, becomes
\begin{equation}\label{e:FP}
\partial_tc+\partial_y\left( c\alpha \int_{-1/2}^{1/2}\text{H}(y-z)(1-c(z,t))\mathrm{d}z\right) =\partial_{yy}(f(c)c).
\end{equation}
In this non-dimensional formulation of the problem, the total mass conservation constraint \eqref{e:total_mass_1} becomes
\begin{equation}\label{total_mass_2}
\int_{-1/2}^{1/2}c(y,t)\mathrm{d}y=1.
\end{equation}
Combining \eqref{e:cell_velocity} and \eqref{e:w_c} and using condition \eqref{total_mass_2}, we obtain the following formula directly relating the velocity and the first moment of the distribution of motors
\begin{equation}\label{e:cell_velocity_2}
V(t)=-\alpha\int_{-1/2}^{1/2}zc(z,t)\mathrm{d}z,
\end{equation}
showing that the cell motion is supported by the global asymmetry of $c$.

When $\alpha=0$, \eqref{e:FP} represents a purely passive system where the motors only diffuse and the solution of \eqref{e:FP} is a homogeneous motor distribution $c\equiv 1$ associated with $V=0$ (and $w\equiv 0$). However, when $\alpha$ becomes larger than the critical value $\alpha_c=\pi^2(f(1)+f'(1))$, where $'$ denotes the derivative, multiple steady states become possible (See Appendix~\ref{sec:appendix_B}) and the question of their local and global stability properties arises. We shall address this question in the following section by exhibiting a  Lyapunov functional that is minimized during the evolution of \eqref{e:no_frict_motil_pb}.

\section{Variational structure} 
We define the Lyapunov functional \cite{frank2005nonlinear,chavanis2015generalized}, $\mathcal{F}=\mathcal{E}-\alpha \mathcal{S}$ where the ``energetic'' and ``entropic'' terms are
$$\mathcal{E}[w]=-\frac{1}{2}\int_{-1/2}^{1/2}w^2\mathrm{d}y \text{ and } \mathcal{S}[c]=-\int_{-1/2}^{1/2}s(c)\mathrm{d}y. $$
\textcolor{black}{Notice that $\mathcal{F}$ is not directly interpretable as a free energy of the system in a classical active gel thermodynamics perspective \cite{PhysRevLett.112.218101}.} In the above formula the entropy per unit volume $s(c)$ is defined in the following way:
$$s''(c)=f'(c)+\frac{f(c)}{c},$$
where we impose that $s(0)=0$ and $s(\infty)=\infty$. As $f$ is a positive and non-decreasing function, these conditions imply the existence of a minimum $s_{\text{min}}\leq 0$ such that $s\geq s_{\text{min}}$. When $f(c)=1$, we recover the Boltzmannian entropy $s(c)=c\log(c)-c$ while for our choice 
\begin{equation}\label{e:steric_choice}
f(c)=1+r c^2,
\end{equation} 
where $r$ is a non-dimensional parameter controlling the strength of the steric hindrance (see Appendix~\ref{sec:appendix_A}), we obtain, 
$$s(c)=rc^3/2+c\log(c)-c.$$ 
For the homogeneous solution, only the entropic term contributes to $\mathcal{F}=\mathcal{F}_0=\alpha (r/2-1)$.

Using \eqref{e:no_frict_motil_pb}, the inequality
$$\partial_t\mathcal{F}=-\alpha\int_{-1/2}^{1/2}\frac{(cw-\partial_y(f(c)c))^2}{c}\mathrm{d}y\leq 0,$$
shows that $\mathcal{F}$ necessarily decays during the dynamics and that $\partial_t\mathcal{F}=0$ implies that $\partial_tc=0$. As using \eqref{e:w_c} we can check that $|w|\leq \alpha$, we also obtain that $\mathcal{F}\geq -(\alpha^2/2-\alpha s_{\text{min}})$ is bounded from below insuring via Lyapunov theory \cite{frank2005nonlinear} that system \eqref{e:no_frict_motil_pb} converges to an equilibrium state. 

The effective energy can be expressed as a functional of $c$ only by using \eqref{e:w_c}, 
$$\mathcal{E}[c]=\frac{\alpha^2}{2} \int_{-1/2}^{1/2}\int_{-1/2}^{1/2}\text{max}(y,z)(1-c(y,t))(1-c(z,t))\mathrm{d}y\mathrm{d}z$$
such that $\mathcal{F}$ is also a functional of $c$ only. Using this expression, we compute the gradient of $\mathcal{F}$ with respect to $c$ 
$$\frac{\delta \mathcal{F}}{\delta c}(y,t)=-\alpha^2\int_{-1/2}^{1/2}\text{max}(y,z)(1-c(z,t))\mathrm{d}z+\alpha s'(c(y,t)).$$
Thus \eqref{e:FP} is equivalent to
$$\partial_tc=\partial_y\left( \frac{c}{\alpha}\partial_y\left( \frac{\delta \mathcal{F}[c]}{\delta c}\right) \right),$$
showing that the dynamics of $c$ is driven by its relaxation to the minimum of the quasi-potential $\mathcal{F}$. The globally stable steady state is therefore the $c_{\text{eq}}(y)$ distribution that minimizes $\mathcal{F}$ under the constraints $\partial_yc_{\text{eq}}(\pm 1/2)=0$ and $\int_{-1/2}^{1/2}c_{\text{eq}}(y)\mathrm{d}y=1$. The local minima of $\mathcal{F}$ are locally stable steady states while maxima and saddle points are unstable steady states \cite{frank2005nonlinear,chavanis2015generalized}. 

\section{Metastable steady-states}

\begin {figure}
\includegraphics [scale=0.6]{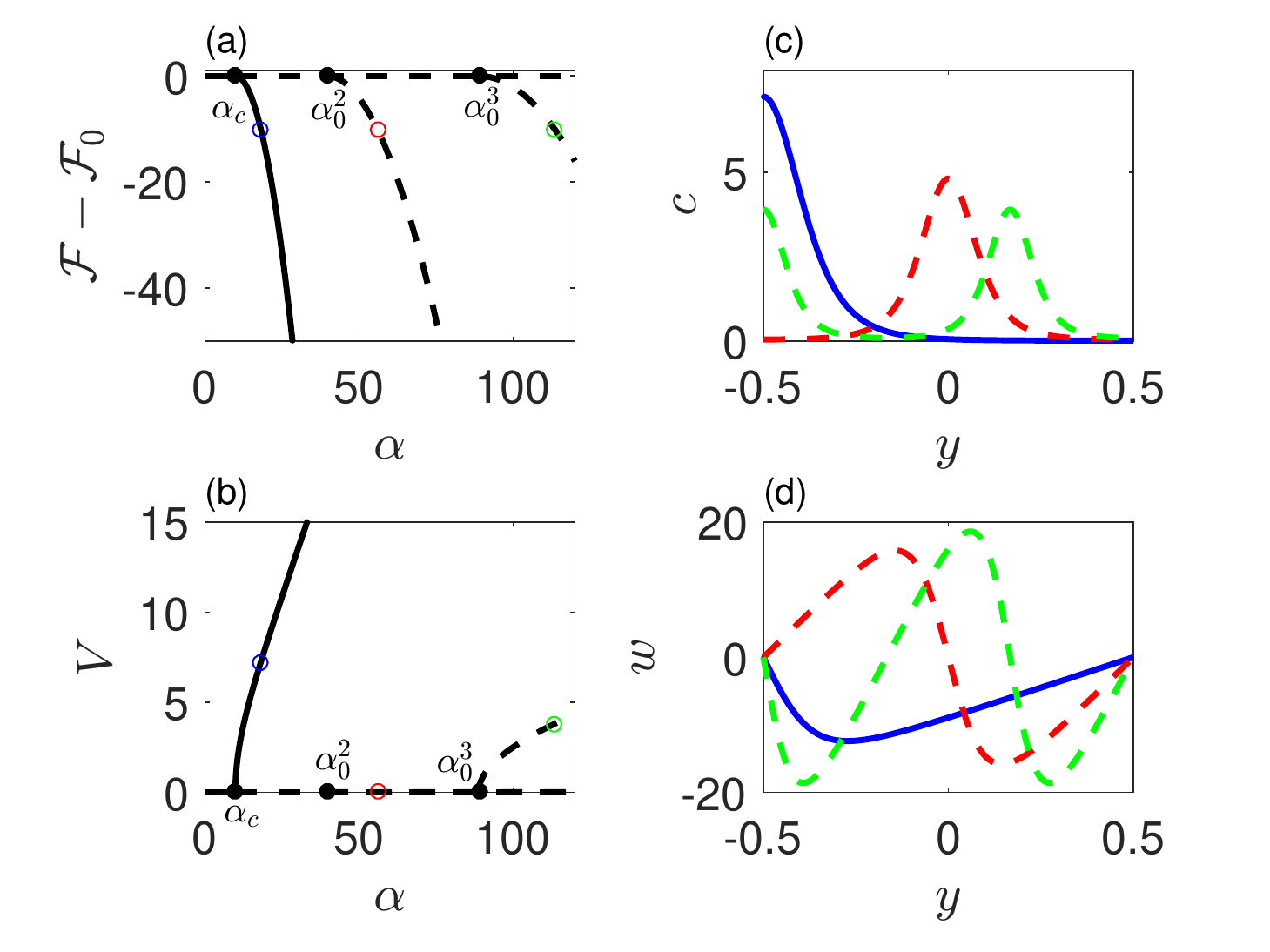}
\caption {Three first bifurcations from the homogeneous state for $r=0$. (a) and (b) are bifurcation diagrams for the quasi-potential and the cell velocity. They have a pitchfork supercritical structure. Black dots localize the bifurcation points. (c) and (d) show the profiles of $c$ and $w$ for some special points labeled with the corresponding colored circles on (a) and (b). Full lines correspond to locally stable branches or solutions while dashed lines are locally unstable}
\label{f:bifr0}
\end {figure}

We begin by characterizing the critical points of $\mathcal{F}$ which correspond to the possible steady states of system \eqref{e:no_frict_motil_pb}. To do so, we implement a continuation method starting from the homogeneous solution at $\alpha=0$ using the  software AUTO \cite{DoeKelKer_ijbc91} and follow into the non-linear regime the bifurcations branching from this state as $\alpha$ increases. The critical values at which these non-trivial solution emerge are given by  $\alpha=\alpha_0^k=(1+3r)k^2\pi^2$, where $k\geq 1$ is an integer (see Appendix~\ref{sec:appendix_B}). The first of these values is  $\alpha_c=\alpha_0^1$. We show the first three branches obtained this way in Fig.~\ref{f:bifr0} for $r=0$. 
As solution measures, we show the values of $\mathcal{F}-\mathcal{F}_0$ and $V$. 
For each solution bifurcating at an odd bifurcation point (i.e. $k$ is odd), there is a symmetric solution with respect to the center of the segment associated with the opposite velocity (see \cite{Recho2015}). The value of the quasi-potential for these two symmetric solutions is the same and we only show the solution leading to a positive velocity in Fig.~\ref{f:bifr0}. Each solution bifurcating at an even bifurcation point (i.e. $k$ is even) has an even symmetry with respect to zero and is thus associated with a zero velocity (see \eqref{e:cell_velocity_2}).  As we show in Fig.~\ref{f:bifr0}, when the bifurcation order increases, the number of patterns in the motor concentration increases. We  check in Appendix~\ref{sec:appendix_C} that, except for the first bifurcation, all the bifurcating solutions are locally unstable. Added to this, the homogeneous solution ceases to be locally stable past the first bifurcation point.

However, the stability status of the first bifurcation branch is interesting. We can analytically show using a normal form expansion (See Appendix~\ref{sec:appendix_B}) that  the bifurcation is pitchfork supercritical if  $r<r_c=(7+\sqrt{57})/12$ or subcritical if $r>r_c$. In the supercritical case, a local stability of the bifurcating branch is found, leading to a simple situation where the cell converges to either a motile or static (homogeneous) state depending whether $\alpha\geq \alpha_c$ or $\alpha\leq \alpha_c$.  The subcritical case is more complex. 
\begin {figure}
\includegraphics [scale=0.6]{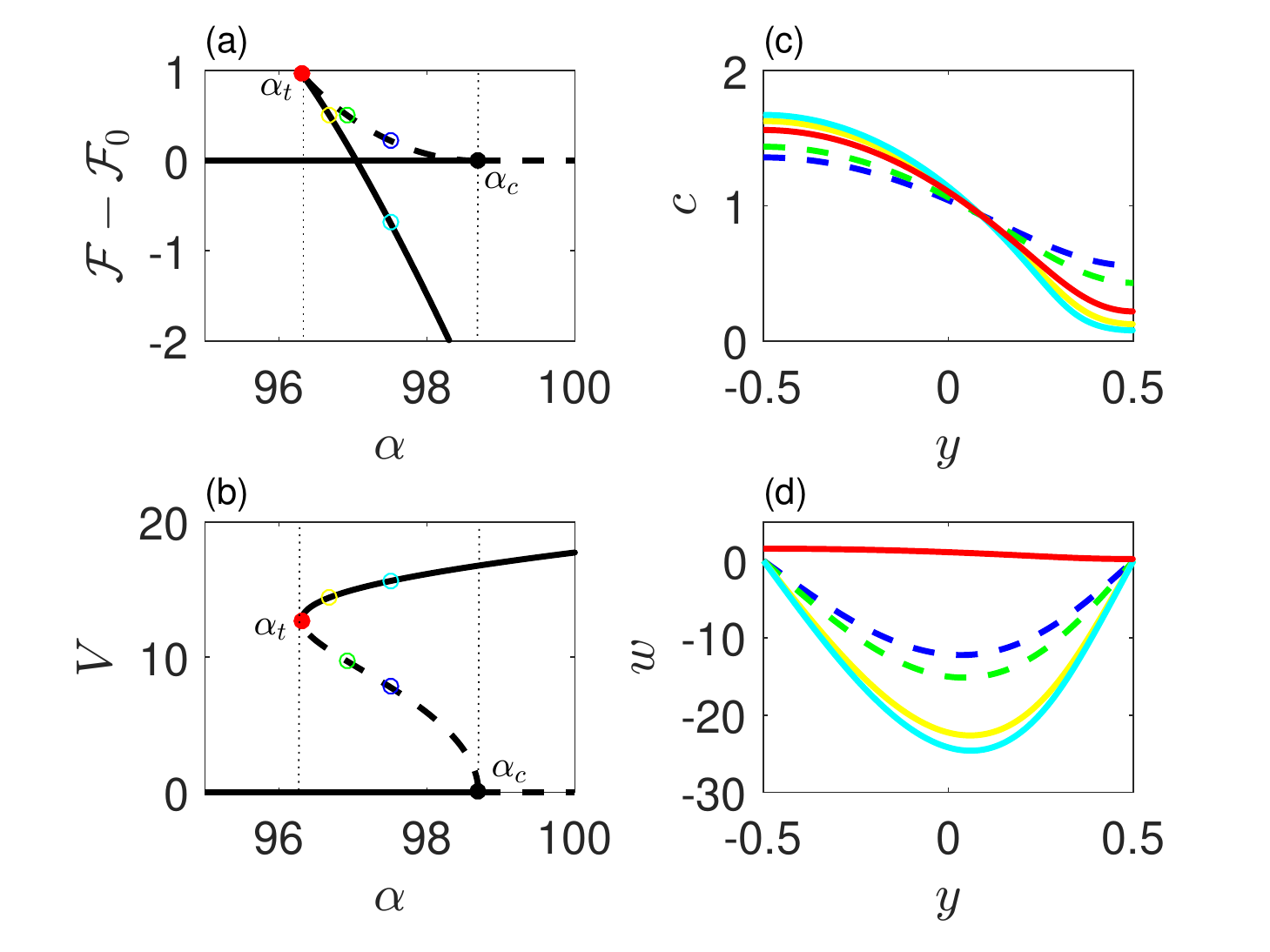}
\caption {Structure of the first bifurcation from the homogeneous state for $r=3$. (a) and (b) are bifurcation diagrams for the quasi-potential and the cell velocity showing the subcritical nature of the bifurcation. The black dot localizes the first bifurcation point and the red dot the turning point. The thin dotted vertical lines represent the domain where both the static and motile configurations are locally stable.  (c) and (d) show the profiles of $c$ and $w$ for some special points labeled with the corresponding colored circles on (a) and (b). Full lines correspond to locally stable branches or solutions while dashed lines are locally unstable.}
\label{f:bifr3}
\end {figure} 
As we illustrate in Fig.~\ref{f:bifr3}, there is a turning point located at $\alpha=\alpha_t\leq \alpha_c$ along the bifurcating branch leading to a fold. We can then again check numerically that solutions before the fold are numerically unstable while solutions after the fold are linearly stable again, although they look qualitatively similar with motors self organizing at the trailing edge of the cell, see Fig.~\ref{f:bifr3}. Thus, there is a choice of parameters ($r>r_c$ and $\alpha \in [\alpha_t,\alpha_c]$) for which the static and  motile configurations can be both locally stable, the globally stable solution being the one corresponding to the minimum of the quasi-potential.
\begin {figure}
\includegraphics [scale=0.5]{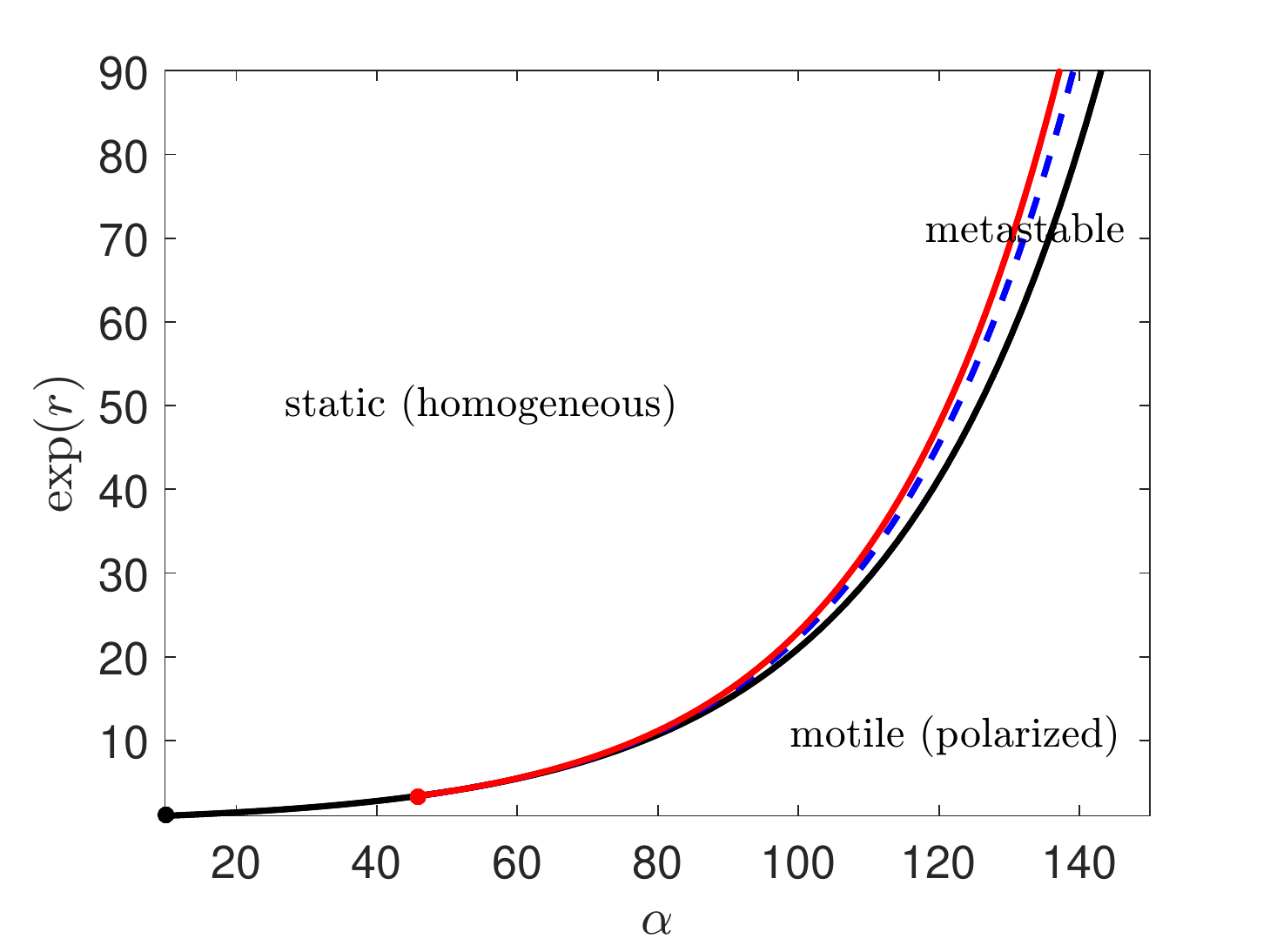}
\caption{Phase diagram in the parameter space $(\alpha,r)$ characterizing the steady state of system \eqref{e:no_frict_motil_pb}. The black line is the locus of the first bifurcation point and the red line the one of the turning point along the first bifurcating branch (when it exists). The blue dashed line represents a ``Maxwell line''. Above this line, the homogeneous solution is the global minimum of the Lyapunov functional $\mathcal{F}$ while below this line, it is the non-trivial polarized solution. We use $\exp(r)$ instead of $r$ to better graphically visualize the separation between the bifurcation, turning point and Maxwell lines.}
\label{f:phase_diag}
\end {figure} 
We show in Fig.~\ref{f:phase_diag} the resulting phase diagram where the motile and static phase are shown as well as the third metastable phase where the two configurations can coexist. In this phase, a ``Maxwell line'' separates the region of parameters space where the motile state is the global minimum of $\mathcal{F}$ and those where it is the static (homogeneous) state. 

This property entails interesting  consequences when the contractility is no longer deterministic but is subjected to small stochastic fluctuations as the cell can switch between the two configurations leading to stop-and-go dynamics.

\section{Stochastic contractility}


To simply illustrate the effect of metastability on the cell dynamics, we consider a source of noise in the model by changing \eqref{e:const_be} into 
$$\sigma=\eta \partial_xv+\chi c+\Sigma_s,$$
where $\Sigma_s(x,t)$ is a small ($|\Sigma_s|\ll \chi c_0$) stochastic spatio-temporal noise. As an example, we take 
$$\partial_t \Sigma_s-\Theta\partial_{xx}\Sigma_s=\dot{W}$$
where $\Theta$ is a diffusion coefficient and $\dot{W}(x,t)$ is a spatio-temporal white noise. Thus $\Sigma_s$ represents small variations of the mechanical stress in the cell skeleton due to some existing random disorder. The non-dimensional model \eqref{e:no_frict_motil_pb} then becomes
\begin{equation}\label{no_frict_motil_pb_noise}
\left\lbrace \begin{array}{c}
\alpha(1-c-\delta\sigma_s)= \partial_yw\\
\partial_tc+\partial_y(cw-\partial_y(f(c)c))=0\\
\partial_t\sigma_s-\theta\partial_{yy}\sigma_s=e\dot{\omega},
\end{array}\right. 
\end{equation}
where the new non-dimensional variables are $\theta=\Theta/D$ that quantifies the spatio-temporal correlation of the noise and $e\ll 1$ that represents the small noise magnitude in the system. $\dot{\omega}$ is a normalized white noise such that, denoting $\langle.\rangle$ the ensemble average,
$$\langle\dot{\omega}(y,t)\rangle=0\text{ and }\langle\dot{\omega}(y,t)\dot{\omega}(y',t')\rangle=\delta(y-y')\delta(t-t').$$
The stochastic stress $\sigma_s=\Sigma_s/(\chi c_0)$ is shifted by 
$$\delta\sigma_s(y,t)=\sigma_s(y,t)-\int_{-1/2}^{1/2}\sigma_s(y',t)\mathrm{d}y'$$
such that it has a zero spatial average.
\begin {figure}
\includegraphics [scale=0.35]{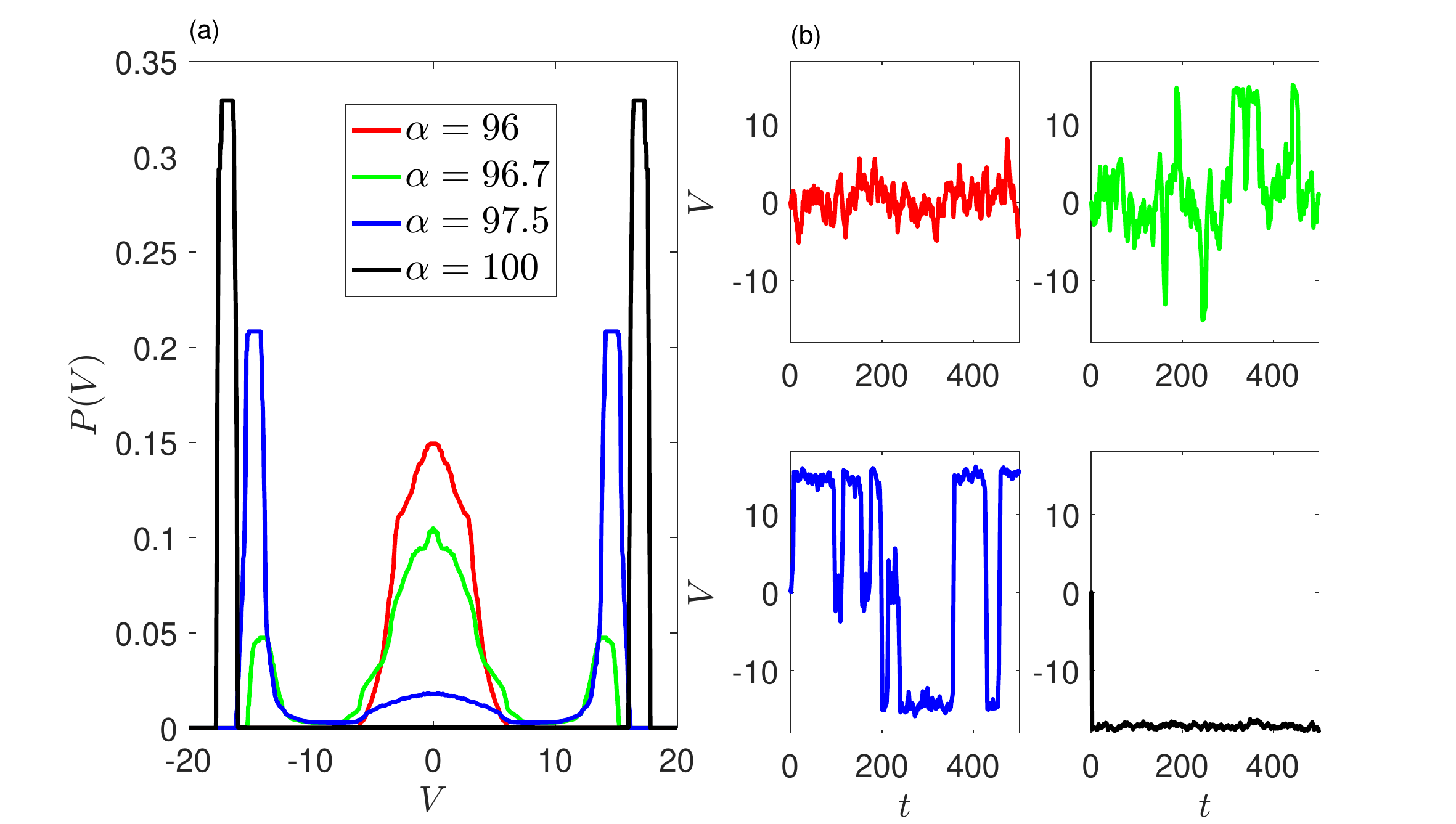}
\caption{Effect of stochastic fluctuations on the cell metastable dynamics defined by system \eqref{no_frict_motil_pb_noise}. (a) Probability densities of the distribution of velocity of a moving cell in four typical cases: in red the static configuration is the only steady state of the deterministic cell dynamics, in green both static and motile states are locally stable but the static state is the global minimum of the quasi-potential, in blue the motile state becomes the global minimum and in black only the motile state is locally stable. (b) shows samples of the velocity dynamics in the four cases.  Parameter $r=3$ and parameters defining the noise are $\Theta=0.01$ and $e=0.001$. The simulations to obtain the probability densities start from the homogeneous distribution and are ran over a non-dimensional time of 1000. The transient state is removed and the distributions are symmetrized with respect to $V=0$ to minimize the computation cost.}
\label{f:stochastic_contract}
\end {figure} 

Next, we chose $r=3$ and numerically simulate \eqref{no_frict_motil_pb_noise} for four values of $\alpha=96$, $96.7$,$97.5$ and $100$. The two central values correspond to a metastable regime, see Fig.~\ref{f:bifr3}, where either the static state or the motile state is the global minimum of the quasi-potential while the other state is a local minimum. We show in Fig.~\ref{f:stochastic_contract}, the typical  dynamics as well as the probability densities of the cell velocities for all four cases. When the static state is the only existing -and stable- steady state of the deterministic system, the velocity is peaked around $V=0$. Then, as we reach the metastable regime, the distribution has three peaks corresponding to a static state and the two symmetric motile configurations. The size of the peaks of the probability density of $V$ depends on which state is the global minimum of $\mathcal{F}$ and the system may feature predominantly fluctuations around the static state with rare motile excursions or, on the contrary, a motile dynamics rarely alternating the sign of the velocity and spending a small duration around the static state. As $\alpha$ increases such that the system leaves the metastable domain, the unstable static state disappears from the velocity distribution.

It is \textcolor{black}{potentially} interesting to interpret these  results at the collective level as metastability can  \textcolor{black}{qualitatively} explain why, in a cell population with the same parameters defining their molecular motors dynamics,  most of the cells may be almost static with only a certain proportion moving at a large velocity or, on the contrary, most cells can be motile and a few of the them static depending which state is the global attractor of the deterministic system.

\section{Conclusions}

We have exhibited one of the simplest model of cell \textcolor{black}{motion} that is independent of its interaction with the substrate as, while they exert vanishingly small traction forces, the molecular motors still produce an internal flow of skeleton that can propel the cell boundary. Such flow has to be coupled with a physical process that insures the recycling of the skeleton building blocks and which is not solved for in this minimalist model. This can be achieved by considering a backflow \cite{loisy2019tractionless} or a chemical  turnover reaction that depolymerizes the skeleton at the back and polymerizes it at the front \cite{putelat2018mechanical}. \textcolor{black}{We show in details in  Appendix~\ref{sec:appendix_0} that the present model can emerge from such perspective.}  This substrate independent \textcolor{black}{motion} mode has a variational structure with a quasi-potential that allows to characterize the local and global stability of its steady states. In particular, we find that there exists a region in the  non-dimensional parameter space where a static and mobile configuration can co-exist in a metastable fashion. In the presence of an additional small  stochastic stress, this leads to the possibility of an intermittent cell dynamics where the static or motile phases of motion dominate depending on which state is the global minimum of the quasi-potential.

\textcolor{black}{It may be interesting to generalize our results outside of the vanishing friction limit where the power of the traction forces is not negligible compared to the internal viscous dissipation. While an intermittent dynamic can still be observed in a certain parameters range in this case, it remains unclear whether it is possible or not to find a quasi-potential that would precisely specify the stability of the steady states. }

\acknowledgements{P.R is thankful to Lev Truskinovsky, Arnaud Millet and Giovanni Cappello for stimulating discussions and references and to Claude Verdier and Jocelyn Etienne for correcting and commenting the manuscript. This work was supported by a CNRS MOMENTUM grant.}

\appendix

\section{Effective diffusion of molecular motors with steric hindrance}\label{sec:appendix_A}

We consider two concentrations  of molecular motors: $c(x,t)$ the concentration of motors that cross-link two fibers of the cytoskeleton (concentration $c$) and  $m(x,t)$ the concentration of motors that are free to diffuse (coefficent $D_m)$ in the cytoplasm \cite{Rubinstein2009}. There is an attachment (rate $k_a$) and detachment (rate $k_d$) dynamics between these two populations that lead to the following coupled system:
\begin{align}\label{e:two_motors}
\partial_tc+\partial_x(cv)=k_a m-k_d c\\
\partial_tm-D_m\partial_{xx}m=k_d c-k_a m.\nonumber
\end{align}    
While we assume that the rate of detachment $k_d$ is fixed, the rate of attachment $k_a=k_a^0g(c)$ decreases with the concentration $c$ because of steric hindrance.  The function $g(c)$ is therefore a positive and decreasing to zero as $c$ becomes large. 

 Assuming that the system remains close to its  chemical equilibrium because the rates are large compared to the transport and diffusion ($k_a,k_d\gg v/L,D/L^2$), we have that
$$m\approx \frac{k_d}{k_a^0} \frac{c}{g(c)}.$$
Plugging this approximation in \eqref{e:two_motors} and assuming that $k_d/k_a^0$ is a small parameter while $D=D_mk_d/k_a^0$ remains finite, we obtain the equation  \eqref{e:motors_dd_1}  by setting that $f(c/c_0)=1/g(c)$ where the scaling parameter $c_0$ is the average concentration of motors that is constant during the dynamics. 

\section{Normal forms of the solutions bifurcating from the homogeneous solution}\label{sec:appendix_B}

The steady states of \eqref{e:no_frict_motil_pb}, for which $\partial_tc=0$ correspond to the solutions of the equation
\begin{equation}\label{e:c_ss}
\partial_y\left( \frac{\partial_y(f(c)c)}{c}\right)+\alpha(c-1)=0
\end{equation}
with Neumann boundary conditions at $y=\pm 1/2$. Eq.~\eqref{e:c_ss} has the homogeneous solution $c\equiv 1$. From this solution, non-trivial solutions bifurcate at specific values of $\alpha$. These bifurcation points and the behavior of the bifurcating solutions can be investigated by plugging a Taylor expansion of $c$ and $\alpha$ in Eq.~\eqref{e:c_ss},
\begin{align}
c(y,t)&=1+\epsilon c_1(y)+\epsilon^2 c_2(y)+\epsilon^3 c_3(y)+...\label{e:Taylor_exp}\\
\alpha&=\alpha_0+\epsilon \alpha_1+\epsilon^2 \alpha_2+\epsilon^3 \alpha_3+...\nonumber
\end{align}
where the root mean square of the $c_i$ is fixed to one and $\epsilon$ is a small parameter.

At first order we find that the operator 
$$(f(1)+f'(1))\partial_{yy}c_1+\alpha_0c_1=0,$$ 
with Neumann boundary conditions becomes degenerate  at the  values of $\alpha_0$ indexed by the integer $k\geq 1$:
$$\alpha_0^k=(f(1)+f'(1))k^2\pi^2.$$
The smallest value of $\alpha_0$ corresponding to $k=1$ is denoted $\alpha_c$. At each $\alpha_0^k$, a solution bifurcates along the two symmetric eigenvectors 
$$c_1^k(y)=\pm\sqrt{2}\cos(\pi k (y+1/2)).$$

At the second order in $\epsilon$, we obtain using the Fredholm alternative that $\alpha_1^k=0$ and 
\onecolumngrid
$$c_2^k(y)=  \frac{c_1^k(y) \sqrt{22 f(1) f'(1)+7 f'(1)^2+4 \left(f(1)-f'(1)\right) f''(1)+7 f(1)^2-2 f''(1)^2}+\sqrt{2} c_1^k(2 y) \left(f(1)-f''(1)-f'(1)\right)}{3 \left(f'(1)+f(1)\right)}$$
\twocolumngrid
Finally, the value of $\alpha_2^k$ fixing the local nature of the bifurcation is classically given by the third order expansion:
\onecolumngrid
$$\alpha_2^k=\frac{\pi ^2 k^2 \left(-4 f''(1)^2-10 f'(1)^2+f(1) \left(3 f^{(3)}(1)+11 f''(1)+8 f'(1)\right)+f'(1) \left(3 f^{(3)}(1)-5 f''(1)\right)+2 f(1)^2\right)}{12 \left(f'(1)+f(1)\right)}$$
\twocolumngrid
Taking the simple form $f(c)=1+rc^2$ where $r$ is a non-dimensional parameter fixing the strength of the  steric hindrance, we obtain
$$\alpha_2^k=\frac{\pi ^2 k^2 \left(-18 r^2+21 r+1\right)}{18 r+6},$$
which is positive for $r<r_c=(7+\sqrt{57})/12$ indicating a super-critical pitchfork bifurcation  while it becomes negative when $r>r_c$ indicating a sub-critical pitchfork bifurcation.

\section{Local stability}\label{sec:appendix_C}

The local (or linear) stability of a certain steady state $c_{\text{eq}}(y)$ is given by the second variation of $\mathcal{F}$ at this point. Based on the expressions of $\mathcal{E}$ and $\mathcal{S}$, we obtain the following quadratic form:
\begin{align}
\delta^2\mathcal{F}[h]=&\frac{\alpha^2}{2}\int_{-1/2}^{1/2}\int_{-1/2}^{1/2}\text{max}(y,z)h(z)h(y)\mathrm{d}y\mathrm{d}z\label{e:second_var}\\
&+\frac{\alpha}{2}\int_{-1/2}^{1/2}s''(c_{\text{eq}}(y))h(y)^2\mathrm{d}y\nonumber.
\end{align}
If $\delta^2\mathcal{F}$ is strongly positive for all test functions $h$ that satisfy the Neumann boundary conditions at $\pm 1/2$ and the constraint
$$\int_{-1/2}^{1/2}h(y)\mathrm{d}y=0,$$
the steady state $c_{\text{eq}}$ is linearly stable. It is unstable otherwise. Such condition is equivalent to checking the positivity of the eigenvalues of the polar form associated to $\delta^2\mathcal{F}$. This leads to the eigenvalue problem 
$$\alpha^2\int_{-1/2}^{1/2}\text{max}(y,z)h(z)\mathrm{d}z+\alpha s''(c_{\text{eq}}(y))h(y)\mathrm{d}y=\mu h(y),$$
where $\mu$ is the eigenvalue and $h$ the eigenvector. Differentiating twice this relation, we obtain the boundary value problem
\begin{equation}\label{e:stab_eigenpb}
\begin{array}{c}
\alpha^2h(y)=\partial_{yy}\left((\mu-\alpha s''(c_{\text{eq}}(y)))h(y) \right) \\
\text{ with } \partial_yh(\pm 1/2)=0. 
\end{array}
\end{equation}
Each eigenvector being defined up to a constant, we additionally impose the normalization
$$\int_{-1/2}^{1/2}h(y)^2\mathrm{d}y=1.$$

The local stability of the homogeneous solution $c_{\text{eq}}(y)\equiv 1$ can be resolved analytically since the solution of \eqref{e:stab_eigenpb} is explicit in this case and we obtain:
$$\mu=\frac{-\alpha^2}{k^2\pi^2}+\alpha(f(1)+f'(1)),$$
where $k\geq 1$ is a positive integer. As a consequence, there exists at least one negative eigenvalue as soon as $\alpha>\alpha_c$ indicating the loss of local stability of the homogeneous solution past the first bifurcation point.

For the non-homogeneous branches, it is not straightforward to solve \eqref{e:stab_eigenpb} and we investigate the local stability properties numerically by using the test function combining the first $Q$ modes
$$h(y)=\sum_{k=1}^{Q}h_kc_1^k(y)$$
in \eqref{e:second_var}. We thus have to test the positivity of the eigenvalues of the symmetric matrix $\delta\mathbb{F}= \delta\mathbb{E}-\alpha\delta\mathbb{S}$ 
with 
$$\delta\mathbb{E}_{i,j}=\frac{\alpha^2}{2}\int_{-1/2}^{1/2}\int_{-1/2}^{1/2}\text{max}(y,z)c_1^i(y)c_1^j(z)\mathrm{d}y\mathrm{d}z=-\frac{\alpha^2\delta_{ij}}{2i^2\pi^2}$$
and
$$\delta\mathbb{S}_{i,j}=-\frac{1}{2}\int_{-1/2}^{1/2}s''(c_{\text{eq}}(y))c_1^i(y)c_1^j(y)\mathrm{d}y$$
and where $\delta_{ij}$ is the Kronecker symbol and $i,j$ are integers in the interval $[1,Q]$.

\section{\textcolor{black}{Model of the skeleton turnover}}\label{sec:appendix_0}

\textcolor{black}{In this section, we expand the model formulation to represent the implicit material turnover of the cell skeleton that is coupled to its retrograde flow. While in the main text, we consider for simplicity only the skeleton and the molecular motors which actuate it, we shall consider here two additional components in the system: a fluid phase (the cytosol in a cell context) that permeates the skeleton meshwork and the skeleton building blocks that are in solution in the permeating fluid phase (such as actin monomers in a cell context).}

\textcolor{black}{Relying on the porous medium active gel theory presented in \cite{deshpande2021chemo} and considering that the volume fraction of fluid is fixed, we can express the mass balance laws of the skeleton, fluid and skeleton building blocks as
\begin{align}
&\partial_t\rho+\partial_x(\rho v)=k_+b-k_-\rho\label{e:m_bal_skel}\\
&\partial_t\rho_f+\partial_x(\rho_f v_f)=0\label{e:m_bal_fluid}\\
&\partial_tb+\partial_x(bv_f-D_b\partial_xb)=k_-\rho-k_+b,\label{e:m_bal_blocks}
\end{align}
where $\rho(x,t)$ is the density of skeleton, $\rho_f(x,t)$ that of the permeating fluid and $b(x,t)$ is the concentration of building blocks in the fluid. Thus, $k_{\pm}$ are the assumed fixed polymerization and depolymerization rates of the skeleton, $v_f(x,t)$ is the fluid velocity and $D_b$ is a diffusion coefficient characterizing the mobility of the monomers with respect to the fluid. As we do not consider any flux of skeleton, water or skeleton building blocks through the cell membrane during the motion, we have that $\partial_tl_{\pm}(t)=v(l_{\pm}(t),t)=v_f(l_{\pm}(t),t)$ and $\partial_xb(l_{\pm}(t),t)=0$.}

\textcolor{black}{The total stress in a representative volume element is
\begin{equation}\label{e:const_porous}
\Sigma=-p_f+\eta\partial_xv+\chi c,
\end{equation} 
where, we have neglected the skeleton compressibility assuming that on a long time scale, it behaves as a viscous fluid and $p_f(x,t)$ is the pressure in the permeating fluid. In the absence of inertia, force balance imposes that 
\begin{equation}\label{e:f_bal_porous}
\partial_x\Sigma=\xi(v-v_s),
\end{equation}
where $\xi$ is a friction coefficient ecompassing both passive friction and the active friction stemming from the  engagement and disengagement of focal adhesions coupling the skeleton and the substrate \cite{tawada1991protein,Sens2013} as introduced in \eqref{e:f_bal_1}.}\textcolor{black}{ To the force balance \eqref{e:f_bal_porous}, following \cite{ambrosi2016mechanics},  we associate the following boundary conditions $\Sigma(l_{\pm}(t),t)=-\gamma (l_+(t)-l_-(t))$ that account for the presence of a membrane tension $\gamma$.} \textcolor{black}{Finally the fluid motion through the skeleton is described by a Darcy law
\begin{equation}\label{e:Darcy}
v_f-v=-\frac{\kappa}{\eta_f}\partial_xp_f,
\end{equation}
where $\kappa$ is the meshwork permeability and $\eta_f$ the fluid viscosity.}

\textcolor{black}{Using the fact that the permeating fluid is incompressible, we obtain from \eqref{e:m_bal_fluid}  that $\partial_xv_f=0$ which, using the associated boundary conditions, leads to \eqref{e:bc_Stefan} of the main text. }\textcolor{black}{In particular, this implies that the length $L=l_+(t)-l_-(t)$ is a constant. }\textcolor{black}{Added to this, it is also considered that the fluid permeation is fast compared to the velocity of the meshwork itself  at our (long) timescale of interest. This can be quantified by the non dimensional number 
$$\frac{\kappa \chi c_0}{\eta_f D}\simeq 4\times 10^3\gg 1 ,$$
where we used the rough estimates derived from experiments on fish keratocytes \cite{Recho2015, deshpande2021chemo}: $\kappa\simeq 2\times 10^{-16}\text{ m}^2 $, $\chi c_0\simeq 10^3\text{ Pa} $, $\eta_f\simeq 2\times 10^{-3}\text{ Pa s}$ and $D \simeq 0.25\times 10^{-13} \text{m}^2s^{-1}$. We then assume that $\partial_xp_f\simeq 0$ (while the product $\kappa\partial_xp_f$ remains undetermined) and $p_f$ is approximately constant in \eqref{e:const_porous} and \eqref{e:f_bal_porous}. Setting $\sigma=\Sigma+p_f$, we thus obtain \eqref{e:f_bal_1} and \eqref{e:const_be} with the associated boundary conditions \eqref{e:f_bal_bc_1}} \textcolor{black}{where the residual stress at the boundaries $\sigma(l_{\pm}(t),t)=-\gamma L+p_f$}.\textcolor{black}{ Along with the dynamical equation for the molecular motors, we therefore recover the model presented in the main text. This model is augmented with the dynamics for the cytoskeleton density \eqref{e:m_bal_skel} and that of its building blocks \eqref{e:m_bal_blocks}. More specifically, using the above formulated assumptions and the non-dimensionalization of the main text, we can couple,
\begin{equation}\label{e:cyto_dyna_porous}
\left\lbrace \begin{array}{c}
\partial_t\rho+\partial_{y}(\rho w)=\nu_+b-\nu_-\rho\\
\partial_tb-d\partial_{yy}b=\nu_-\rho-\nu_+b
\end{array}\right. 
\end{equation}
to our model system \eqref{e:no_frict_motil_pb}. In \eqref{e:cyto_dyna_porous}, we have kept the same notations for the densities rescaled by the constant $\rho_f$: $\rho:=\rho/\rho_f$ and $b:=b/\rho_f$ and used the non-dimensional quantities $d=D_b/D$ and $\nu_{\pm}=L^2k_{\pm}/D$. Once $w$ is solved for in  \eqref{e:no_frict_motil_pb}, we can solve the coupled drift-diffusion equation determining $\rho$ and $b$ in  \eqref{e:cyto_dyna_porous}. In particular, the cytoskeleton building blocks diffuse in the cytoplasm and are polymerized and depolymerized into the meshwork according to a first order kinetic.
}

\bibliographystyle{apsrev4-1}
\bibliography{zero_friction}

\end{document}